# Newton's Hypotheses on the Structure of Matter

Philip Yock
Department of Physics (ret.), University of Auckland, Auckland, New Zealand
p.yock@auckland.ac.nz

**Abstract.** Isaac Newton's book 'Opticks' from the 18$^{th}$ century includes several hypotheses on the structure of matter. Most of the hypotheses were confirmed during the 19$^{th}$ and 20$^{th}$ centuries at the scale of atoms and molecules. Conflicts appear however with today's Standard Model of particle physics. The confirmations at atomic and molecular levels, and the conflicts at the subatomic level, are described and discussed here. Various precursors to the Standard Model for which the conflicts are less severe are also described. These date back to Yukawa's meson model of 1935, and they are subject to other uncertainties. Observations with electron-proton colliders that are under consideration by CERN could clarify the situation.

1. **Introduction**

Isaac Newton's treatise entitled 'Opticks', which was published in the early 1700s in several editions, included interesting and sometimes provocative hypotheses on the structure of matter (Newton 1730). The hypotheses were expressed quite briefly, and appeared in the closing pages of Opticks interspersed amongst items on other topics. They report a self-consistent but speculative and incomplete viewpoint of the basic properties of matter as envisaged by Newton. Science conducted in the following three hundred years can be used to test his predictions, and perhaps glean further information from the hypotheses that might be useful today.

The dominant thought contained within the hypotheses appears to be Newton's belief in a clustered but ordered structure of matter, commencing with the smallest and most tightly bound clusters, and ending with the largest and most weakly bound ones. Today this pattern can be identified with quarks, nucleons and nuclei at smaller scales, and with atoms, molecules and macromolecules at larger scales.

When the full range of Newton's hypotheses is examined one finds support for his ideas when applied to atoms and molecules, but puzzling disagreements or refutations when applied to quarks and nucleons. This state of affairs is discussed here.

One difficulty that is encountered with the physics of quarks and nucleons is the concept of quark confinement. This does not allow unique conclusions to be drawn with certainty. For this reason a range of theories is considered here to test Newton's hypotheses. Precursors to today's Standard Model of particle physics are considered that extend back to Yukawa's theory of mesons (Yukawa 1935).

It is found that in some respects the precursors fare better with Newton's hypotheses than does the Standard Model. The idea is raised that they might provide useful pointers towards an improved model of particle physics. An experimental programme that is under consideration at CERN could achieve a comparable result. This is described here.

2. **Newton's hypothesis of the clustered structure of matter**

Newton summarized his vision of a clustered structure of matter with the following words (Newton 1730, pp. 394 - 395) that were reproduced previously on at least two occasions in the physics literature (Weinberg 2015, Yock 1970):-

> *'There are therefore Agents in Nature able to make the Particles of Bodies stick together by very strong Attractions. And it is the Business of experimental Philosophy to find them out. Now the smallest Particles of Matter may cohere by the strongest Attractions, and compose bigger Particles of weaker Virtue; and many of these may cohere and compose bigger Particles whose Virtue is still weaker, and so on for divers Successions, until the Progression end in the biggest Particles on which the Operations in Chymistry, and the Colours of natural Bodies depend, and which by cohering compose Bodies of a sensible Magnitude.'*

Nowadays the clusters from smallest to largest can be identified as nucleons, nuclei, atoms, molecules and macromolecules, and the constituents of the clusters as quarks, nucleons, electrons, atoms and molecules according to present theory. It is also easy to identify the binding agent for nucleons and nuclei as the strong nuclear force, and the binding agent for atoms and molecules as electromagnetism. Fig. 1 below shows the familiar structure of the lithium atom as a representative example.

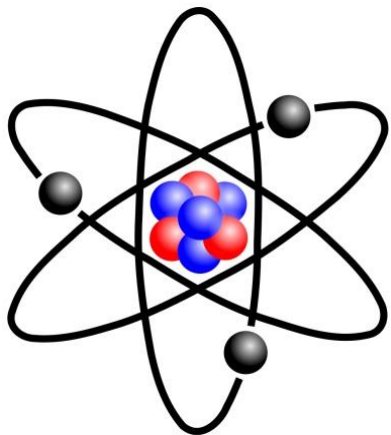

*Fig. 1. Schematic view[1] of a lithium atom, with protons shown in red, neutrons in blue and electrons in black. According to current theory (see §2.1), neutrons and protons are comprised of quarks which are permanently trapped or confined within them. Quarks are therefore unobservable as isolated entities.*

The well-known contributions of Dalton, Mendeleev, Thomson, Rutherford, Chadwick, Schrodinger, Fermi, Dirac, Yukawa, Gell-Mann and others that led to this synthesis during the 19th and 20th centuries are consistent, in a general sense, with Newton's expectation. In what follows, more detailed hypotheses by Newton, including those that question current thinking, are reproduced and discussed.

Needless to say, the advances by Dalton and others were not made by direct visual observation. The resolution of the optical microscope falls short by a factor of order 10,000 of that required to see the atoms that were implied by Dalton's work. It is not possible to see individual atoms, let alone the constituents of the atom. Indirect methods are required.

### 2.1 Hypotheses of self-similarity and simplicity

The severest difficulty posed by Newton to current thinking probably arises in the following extract from Opticks (Newton 1730, p. 398):-

> *'Nature will be very conformable to her self and very simple, performing all the great Motions of the heavenly Bodies by the Attraction of Gravity which intercedes those Bodies, and almost all the small ones of their Particles by some other attractive and repelling Powers which intercede the Particles'.*

---
[1] Courtesy Wikipedia  http://periodictableofelements.wikia.com/wiki/File:Stylised_Lithium_Atom.png

The hypotheses of conformability (i.e. self-similarity) and simplicity may be said to be satisfied in a general sense over a huge range of scales. Electrons orbit nuclei as planets orbit stars, and as stars orbit galaxies. In all cases the causative force is a simple inverse square law. The gravitational forces that drive the heavenly bodies, to use Newton's expression, are universally attractive, whereas the Coulomb forces that drive the components of atoms and molecules are both attractive and repulsive, as Newton predicted. Nuclei are composed of nucleons in contact, and molecules are composed of atoms in contact. The forces that form nuclei and molecules are residuals of the forces that form nucleons and atoms. The various layers of matter are indeed self-similar in a general sense.

Considering the fact that Newton preceded Coulomb, and that no detailed knowledge of atoms or molecules (as opposed to light) was available to him, one can only admire the extraordinary range of applicability of his hypotheses, extending over more than 30 orders of magnitude from atoms to galaxies. Of course, atomic and molecular orbitals require modern quantum theory to be properly understood, but even this fact can be seen as unsurprising from the point of view of Newton's hypotheses (see §2.2 and §2.3 below).

However, when one attempts to test Newton's hypotheses of self-similarity and simplicity at the subatomic level, apparent hurdles appear. One encounters dissimilarity rather than similarity, and complexity rather than simplicity.

According to the current Standard Model of particle physics, sub-atomic particles such as the neutron and the proton are comprised of elementary particles termed 'quarks' which:-

- possess fractional electric charges, either +⅔ or -⅓ in units of the proton's charge
- possess 'colour charges' that may be red, green or blue (see § 3.1 below)
- are confined within neutrons, protons and similar particles in combinations that possess no net fractional or colour charge, thus rendering the fractional and colour charges of quarks effectively unobservable.

The fractional and colour charges of quarks, and their inability to be extracted and examined in isolation, differentiate quarks from all other entities in science. They cannot be said to comprise a self-similar layer in the overall scheme of things[2].

In addition, proof of the confinement of quarks has not been forthcoming in the 45 years that have elapsed since the Standard Model was formulated (Fritzsch et al. 1973). This is true despite a reward of $1M being on offer since 2000 for a (partial) proof[3]. Confinement as presently understood cannot be described as simple.

Weinberg's recently made statement (Weinberg 2015) '*as science progressed after Newton a remarkable picture began to take shape: it turned out that the world is governed by natural laws far simpler and more unified than had been imagined in Newton's time*' may therefore be regarded as unconfirmed at the present time.

The above failures of the Standard Model to satisfy Newton's hypotheses of self-similarity and simplicity are clearly related. The very particles that possess anomalous charges are just the ones that cannot be isolated. The possibility arises that these failures may pinpoint a fruitful area of study for the purpose of better understanding the structure of matter. This is elaborated upon in later sections of this essay.

### 2.2 Hypothesis on attractive forces

---

[2] Gell-Mann (2009) notes, however, that all interactions included in the Standard Model are gauge interactions.
[3] See Millenium Problems at http://www.claymath.org/millennium-problems/millennium-prize-problems

The third quotation of Newton to be cited here expresses his thoughts on the nature of attractive forces (Newton 1730, p. 376):-

*'What I call Attraction may be perform'd by impulse, or by some other means unknown to me. I use that Word here to signify only in general any Force by which Bodies tend towards one another, whatsoever be the Cause'.*

Nowadays it is known that the impulsive interactions that occur in the phenomenon of particle exchange between pairs of particles, as shown in Fig. 2 below, can give rise to both attractive and repulsive forces according to the quantum field theories that were developed in the 20th century. Classically, one would expect repulsive forces only to arise, but quantum uncertainty allows both attractive and repulsive forces to be generated, depending on the spin of the exchanged particle (see, e.g., Deser 2005).

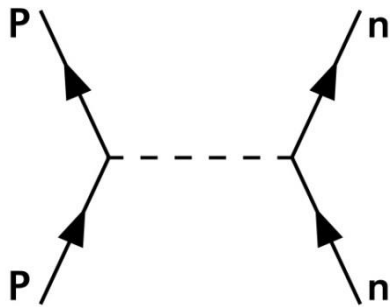

*Fig. 2. The process of π-meson exchange, shown by the dashed line, between a proton and a neutron yields an attractive force according to quantum field theory (Yukawa 1935).*

Newton's assumption of attractive forces arising through an impulsive mechanism via means (viz. special relativity and quantum uncertainty) that were unknown to him displayed remarkable, if not unique, foresight. The development of quantum mechanics required 200 years post Newton, but his thoughts 300 years ago appear to have been extraordinarily well directed.

### 2.3 Hypothesis of elementary particles of finite size

Newton expressed thoughts on elementary particles of finite size, which are included here largely for sake of completeness, as follows (Newton 1730, p. 400):-

*'All these things being consider'd, it seems probable to me, that God in the Beginning form'd Matter in solid, massy, hard, impenetrable, moveable Particles, of such Sizes and Figures, and with such other Properties, and in such Proportion to Space, as most conduced to the End for which he form'd them; and that these primitive Particles being Solids, are incomparably harder than any porous Bodies compounded of them; even so very hard, as never to wear or break in pieces; no ordinary Power being able to divide what God himself made one in the first Creation.'*

Majority opinion nowadays would surely dismiss the above, and Newton does indeed qualify his opinion here as being merely 'probable'. However, if one considers the situation in light of Newton's three laws of motion, even with Coulomb and gravitational interactions included phenomenologically, one can perhaps understand the origin of the above hypothesis. Given only the above laws, there may be no mechanism to prevent the collapse of material acting under the influence of attractive forces to infinite density other than an assumption of elementary particles of finite size.

It is of course quantum mechanics that yields grounds states of finite size for point-like constituents, such as the finite radius of the ground state of hydrogen, but this result was not available to Newton. Once again, Newton appears to have displayed remarkable foresight.

### 2.4 Hypothesis on purpose

Newton expressed an opinion on the presence of purpose and other such characteristics in science as follows (Newton 1730, p. 369):-

*'What is there in places almost empty of Matter, and whence is it that the Sun and Planets gravitate towards one another, without dense Matter between them? Whence is it that Nature doth nothing in vain; and whence arises all that Order and Beauty which we see in the World? To what end are Comets, and whence is it that Planets move all one and the same way in Orbs concentrick, while Comets move all manner of ways in Orbs very excentrick; and what hinders the fix'd Stars from falling upon one another?'*

Here he professes to being of the persuasion that nature does nothing in vain, another stand that currently receives scant attention. He also re-raises the question of the nature of attractive forces and action at a distance, makes clear his lack of knowledge of the rotational motion of the stars of the galaxy, and states his opinion on the beauty and order of nature.

Focussing on the first of these matters, we note that life as we know it requires the presence of many of the chemical elements of the periodic table, but that this does not, in itself, prove the case for purpose.

However, a case for the converse appears to be posed by the quarks of the Standard Model. According to the model, there are six varieties of quarks, known respectively by the rather inelegant names 'up', 'down', 'charmed', 'strange', 'top' and 'bottom'. Matter containing any of the last four types of quarks is highly unstable according to the model. Normal matter therefore contains up and down quarks only, and the existence of the charmed, strange, top and bottom quarks appears to be superfluous.

This begs the question, why should the latter four types of quarks exist? Similar questions were first raised by Isidor Rabi in the 1930s (unpublished), and more recently by Feynman (1985), Weinberg (2015) and others.

In summary, the evidence on purpose as envisaged by Newton appears consistent with observation at the atomic and molecular scales, but inconsistent at the level of quarks.

### 2.5 Hypothesis on the origin of the universe

Newton commented in Opticks on the origin of the universe as follows (Newton 1730, p. 369):-

*'Whereas the main Business of natural Philosophy is to argue from Phænomena without feigning Hypotheses, and to deduce Causes from Effects, till we come to the very first Cause, which certainly is not mechanical; and not only to unfold the Mechanism of the World, but chiefly to resolve these and such like Questions.'*

Here Newton appears to express the opinion that the origin of the universe and its subsequent evolution are different matters. The hypothesis as such goes beyond the realm of physics, and it is included here for sake of completeness without further comment.

### 2.6 Newton's hypotheses in the light of modern physics

Despite Newton having no detailed knowledge whatsoever of the structure of matter, his hypotheses are seen to be applicable at the level of atoms and molecules with remarkable acuity. On the other hand, striking conflicts appear at the subatomic scale as understood by today's Standard Model of particle physics. The latter model does not display the qualities of self-similarity, simplicity or purpose that Newton anticipated.

The contrasting nature of these conclusions is puzzling. How could Newton have been so right with atoms and molecules, yet so wrong with particles?

The Standard Model effectively treats quarks as free and yet confined particles, and this could clearly be a contributory source of conflict between Newton's hypotheses and the Standard Model, especially because Newton is unlikely to have entertained such a possibility. The precise mechanism that prevents free quarks from escaping from high-energy collisions of normal particles, such as those that occur between protons in the Large Hadron Collider (LHC), is not normally included in fits to data[4]. It is not clear that the Standard Model is tested in these analyses.

Rather than attempting to understand the origins of the various conflicts between the Standard Model and Newton's hypotheses, in what follows an alternative procedure is followed. The possibility that Newton's hypotheses could have been essentially correct is examined by considering alternatives to the Standard Model that respect his hypotheses.

### 3. Precursors to the Standard Model that respect Newton's hypotheses

In what follows, we continue the foregoing historical approach by considering precursors to the Standard Model only, all of which were proposed about 50 years ago, or more.

### 3.1 Yukawa model

By 1935 it had been found in table-top experiments that atomic nuclei from the lightest to the heaviest in the periodic table could, with high confidence, be visualized as closely-packed systems of protons and neutrons as shown in Fig. 1 above, where the radii of the neutrons and protons were about $1.3 \times 10^{-15}$ m (Pollard 1935, Evans 1955). This result was subsequently confirmed and refined in electron scattering experiments conducted at Stanford University in the 1950s which yielded radii of $(1.07 \pm 0.02) \times 10^{-15}$ m for neutrons and protons (Hahn et al 1956).

The above results implied the presence of an attractive force between nucleons with a very short range of about $2 \times 10^{-15}$ m, and also a repulsive force of even shorter range to maintain the constant separation between neighbouring nucleons that was observed throughout the periodic table in the above experiments. The combination of attractive and repulsive forces could yield atomic nuclei of approximately constant density from hydrogen to lead, and beyond, as observed. Contemporaneous measurements of the binding energies of atomic nuclei implied that the combined force was strong (Ashton 1936).

In 1935 Yukawa noted that an attractive force between nucleons with range $r$ could be produced by the exchange of a spinless[5] particle of mass $\hbar/rc$. This was consistent with Newton's assumption of attractive forces arising by an impulsive mechanism (see §2.2 above). Yukawa predicted a mass for the exchanged particle of about 200 $m_e$, and this was confirmed in 1947 by the discovery of the pi-meson in the cosmic radiation with a mass of about this value (Lattes et al. 1947). Later observations of the pi-meson confirmed that it interacted strongly as required by the theory, that it had a spin of zero as required, and that its mass was ≈ 270 $m_e$.

Subsequently, other strongly interacting mesons of spin one and greater mass were found that could supply a repulsive force of shorter range (see, e.g., Maglic et al. 1961), and eventually a full theory of nuclear interactions that was well satisfied at MeV energies was developed along the lines initiated by Yukawa (see, e.g., Machleidt 1989).

---

[4] See the process of so-called 'hadronization' described at https://en.wikipedia.org/wiki/Hadronization
[5] The spin of a particle is its intrinsic angular momentum measured in units of Planck's constant ℏ.

In retrospect, the prediction and discovery of the pi-meson appears to have been one of the major successes of quantum field theory, and indeed of theoretical physics. Yet, despite this, the Yukawa mechanism has been abandoned in recent years by the particle physics community in favour of an alternative theory known as quantum chromodynamics or QCD[6]. The latter theory is a basic component of the Standard Model (Fritzsch et al. 1973) although it was formulated nine years after the original proposal of quarks was made (Gell-Mann 1964). The prime function of QCD in the model is to glue quarks together within protons, neutrons and similar sub-atomic particles.

As stated in §2.1 above, QCD posits that quarks are dually charged, carrying both fractional electric charge and also a second type of charge known as 'colour charge'. The theory assumes the presence of three types of colour charge, which are termed 'red', 'green' and 'blue'. Combinations of quarks that are neutral with respect to the colour charge are assumed to bind together, in analogy to the electrical neutrality of atoms, but neutrality with respect to colour is assumed to be achieved by combining equal mixtures of quarks of the three primary colours in analogy to the three-colour model of white light, or by combining quarks with antiquarks.

QCD is an aesthetically pleasing theory, and it accounts well for the observed pattern of hadrons, rather like Mendeleev's table of the chemical elements. But there are exceptions to the classification scheme. Most obviously, particles have been detected which require four or five quarks according to the model (Cho 2016). These were not expected. Conversely, 'glueballs' were expected but have not been found[7]. In addition, an ongoing effort to reproduce the above described nucleon-nucleon force by QCD has met with questionable success only to date (Ishii et al. 2007, Doi et al. 2017).

Other disagreements occur with QCD where the Yukawa model succeeds. Some years ago it was realised that the Yukawa theory made detailed and successful predictions of the polarizations of particles produced in interactions of particles at GeV energies via the Treiman-Yang test (Treiman and Yang 1962, Gottfried and Jackson 1964, Yock and Gordon 1966). The Treiman-Yang mechanism does not, however, carry over to QCD, and the observed polarizations would appear not to be derivable in QCD.

In addition, interactions with so-called 'rapidity gaps' have been observed more recently in which a high-energy electron interacts with a proton as if it was a bound state of a neutron and a positive pi-meson. Such interactions are expected in a generalized version of the Yukawa model (Yock 2002) but not in QCD (Derrick et al. 1996, Lu et al. 2000).

We conclude that the Yukawa model correctly predicts the radii of nucleons and nuclei, the general nature of nuclear forces and interactions at MeV energies, various detailed observations at GeV and higher energies that cannot be derived in QCD, and that it is fully consistent with Newton's hypotheses. Despite these successes it is, however, incomplete, as described below. At best, it can only be a step in the right direction.

### 3.2 Generalized Yukawa model

The incompleteness of the Yukawa model became apparent soon after the discovery of the pi-meson when a new class of particles was found in the cosmic radiation, now known as the strange particles, that was entirely unexpected (see, e.g., Butler and Rochester 1947). The first attempt to understand the burgeoning number of particles was mounted by Fermi and Yang.

---

[6] Newton might have approved of the title.
[7] Glueballs are predicted particles consisting entirely of quanta of the colour force field of QCD.

8These authors proposed in 1949 that the pi-meson is a bound state of a nucleon and an antinucleon with large binding energy (Fermi and Yang 1949). They extended this concept by modelling the nucleon as a 'bare' nucleon surrounded by a cloud of bare pi-mesons - the so-called 'pion cloud' - as Yukawa had done in 1935. Fermi and Yang found the size of the pion cloud in their model to be an order of magnitude larger than the size of a bare pi-meson.

It is a small conceptual step from the Fermi-Yang model to consider the generalised Yukawa model depicted below in which both the bare nucleon and the bare pi-meson are assumed to be composed of tightly bound and closely spaced constituents.

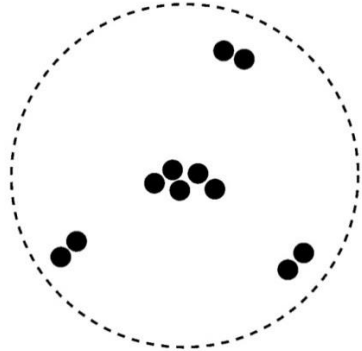

*Fig. 3. Schematic depiction from Yock (2002) of the proton in a generalized Yukawa model in which bare mesons and the bare nucleon are composed of tightly bound constituents. The radius of the proton is ~ $10^{-15}$ m but the spacing between constituents comprising bare particles may be much less (see §4).*

The above extension of the Yukawa model maintains Newton's vision of a clustered structure of matter, with clusters of two scales appearing within the nucleon. The two scales of length provide a means for combining the successful features of the Yukawa model and the quark concept in a coherent fashion. Two attempts to follow this route are described below.

### 3.3 Dyon model

In 1969 Schwinger assumed that hadrons[8] are composed of particles he termed 'dyons' which carried spin ½ and dual charges - fractional electric charges and large magnetic charges (Schwinger 1969). The scale of magnetic charge was determined by Dirac's quantization procedure for charge (Dirac 1931). Strong attraction was assumed to occur between north and south poles, and bound states were assumed to form with large binding energies. The model thus included features of the generalized Yukawa model. It was also consistent with Newton's hypotheses in the sense that Newton included magnetism amongst his candidate binding mechanisms for bound states (Newton 1730, p. 376):-

*'For we must learn from the Phænomena of Nature what Bodies attract one another, and what are the Laws and Properties of the Attraction, before we enquire the Cause by which the Attraction is perform'd. The Attractions of Gravity, Magnetism, and Electricity, reach to very sensible distances, and so have been observed by vulgar Eyes, and there may be others which reach to so small distances as hitherto escape Observation; and perhaps electrical Attraction may reach to such small distances, even without being excited by Friction.'*

---

[8] Hadron is the generic term for particles such as the proton, neutron and pi-meson that interact strongly with one another when they come into contact.

The above excerpt from Opticks was reproduced previously (Gell-Mann 2009, Weinberg 2015). Besides raising the possibility of binding by magnetism, it emphasizes Newton's belief in the crucial importance of observation in science, a belief he practiced.

Despite being consistent with the models noted above, the dyon model required a large electric dipole moment for the neutron, and this is not observed (Baker et al. 2006). The model is therefore included here for completeness, but not examined further.

### 3.4 Subnucleon model

In the same year that Schwinger proposed the model of dyons, the author proposed a comparable model of spin ½ 'subnucleons' as the constituents of nucleons and other hadrons (Yock 1969). The model arose out of an attempt to construct a finite version of quantum electrodynamics (QED), the quantum field theory of photons, electrons and positrons. The aim was to avoid the well-known renormalization divergences that occur in QED (Feynman 1949, Schwinger 1948, Tomonaga 1948, Dyson 1949).

The divergences of QED appear in calculations of the un-renormalized[9] values of the charge and mass of the electron, and elsewhere, and they appear to render the mathematics of QED self-inconsistent (Dirac 1958, Feynman 1985). Little progress was made by the author on the divergences beyond the proposal that the problems they presented might be alleviated or solved if a new class of particles existed that would be akin to the electron but possess high electric charges and high masses (Yock 1969).

The above result was based on a prior effort by Kenneth Johnson and others to construct a finite version of QED (Johnson et al. 1967, Gell-Mann and Low 1954). Their work was generalized to formulate a speculative and incomplete theory of hadron constituents, termed subnucleons, with the afore-mentioned high charges and masses. The high electric charges of subnucleons were utilised to form bound states of subnucleons and antisubnucleons comparable to the bound states of north and south poles in Schwinger's theory of dyons, and comparable to the colour neutral states of the Standard Model that followed in 1973 (Yock 2016).

Fractional electric charges were not included in the subnucleon theory, and electric dipole moments were not predicted. In an attempt to construct a realistic theory, the above-described generalized Yukawa model was followed (Yock 2002). In addition, Newton's hypothesis of electrical attraction reaching to small distances (§3.3) was followed. The model incorporated maximal self-similarity in the sense that all binding, from that of subnucleons in nucleons to molecules in macromolecules, was assumed to be electrical.

Only the barest possible outline of a possible theory was constructed along the above lines. At best it raised the possibility of constructing a finite theory of particle physics following Newton's hypotheses. It is not known by the author if completion of the theory to include all particles and all interactions, including the Higgs, is possible.

### 4. **Observations with electron-proton colliders**

Although the model of the nucleon described above and shown in Fig. 3 is manifestly consistent with Newton's hypothesis of a clustered structure of matter, Newton would be the first to request observational evidence. For this purpose, electron-proton colliders probably offer the best prospects. Just as clear information on the structure of the atomic nucleus was provided by electron scattering (§3.1), comparable processes could be used to seek evidence for the two-scale model of the nucleon depicted in Fig. 3.

---

[9] The un-renormalized values of the electron's charge and mass, $e_0$ and $m_0$, are those that are assumed in the initial formulation of the theory. These differ from the observed values, $e$ and $m$, as a result of the interactions the electron undergoes. In principle, the values of $e_0$ and $m_0$ can be calculated from the observed values, $e$ and $m$, but in practice the calculations are beset with divergences and the values of $e_0$ and $m_0$ are not determined.

Some progress has already been made. A search for particles with charges from 6 to 17 times the charge of the proton was already conducted with the LHC at CERN. This yielded a null result (Aad et al. 2011), implying masses $\geq 1$ TeV/c$^2$ for free subnucleons with charges in the above range, if they exist. Such masses are admittedly extremely high, but nonetheless smaller than the infinite masses that were originally proposed for quarks (Gell-Mann 1964). Masses $\geq 1$ TeV/c$^2$ suggest spacings within the proton between subnucleons in bare nucleons and mesons $\leq 10^{-19}$ m. Evidence for substructure in the proton at this level could be sought with electron-proton colliders.

The resolution achieved in electron-proton interactions is given by $\hbar/Q$ where $Q$ denotes the magnitude of four-momentum transferred by the electron in the scattering process. To achieve a resolution ~ $10^{-19}$ m, momentum transfers ~ 1,000 GeV/c are required.

Momentum transfers of a few × 100 GeV/c were already observed with the HERA electron-proton collider at Hamburg during the 1990s and 2000s. Initial results by two collaborations with a modest integrated luminosity[10] < 0.1 fb$^{-1}$ appeared to indicate threshold behaviour consistent with highly charged constituents beginning to be resolved (Adlof et al. 1997, Breitweg et al. 1997) but the results were subject to the statistical uncertainties of small event numbers. Final results at 10 × higher luminosity were ambiguous (Abramowicz et al. 2016).

It would be of interest to conduct further measurements at higher momentum transfers to probe smaller scales unequivocally. Two machines are already under consideration at CERN that could achieve this goal. They are the Large Hadron Electron Collider or LHeC (Aabdelleira Fernandez et al. 2012), and the Very High Energy Electron Proton Collider or VHEeP (Caldwell and Wing, 2016).

The LHeC has been planned to collide 60 GeV electrons from a purpose-built electron accelerator with 7 TeV protons from the LHC to yield peak values of $Q$ ~ 1000 GeV/c at high luminosities. This would probe substructure at scales down to $10^{-19}$ m.

The VHEeP collider offers the potential advantage of finer resolution. It would collide 3 TeV electrons with 7 TeV protons, where the 3 TeV electrons would be produced by the novel proton-driven plasma wakefield acceleration technique (Caldwell et al. 2009) and the 7 TeV protons would be supplied by the LHC.

The maximum possible transverse momentum $Q$ of the VHEeP collider would be considerably higher than those of either HERA or the LHeC. However, the integrated luminosity of the VHEeP is currently expected to be only ~ 10 - 100 pb$^{-1}$. Remarkably, this could suffice to detect events with $Q$ up to ~ 1000 GeV/c. This follows because the event rate scales as $\alpha^{-1}(Q^2)^{-2}L$ where $\alpha$ = 1/137 if highly charged subnucleons with fine structure constant of order unity are resolved or partially resolved, and where $L$ denotes luminosity. The VHEeP could therefore probe scales in the proton ~ $10^{-19}$ m.

5. **Conclusions**

An attempt has been made here to reproduce and discuss the main elements of Newton's hypotheses from Opticks on the structure of matter. Needless to say, responsibility for any errors or omissions in the selection of items for reproduction, or in their interpretations, lies entirely with the author.

Those words having been said, it is the opinion of the author that the main elements contained in Newton's hypotheses can be fairly said to have displayed extraordinary foresight, and to have been beautifully confirmed by subsequent events at the level of atoms and molecules.

---

[10] 'Integrated luminosity' is the ratio of the number of interactions observed in a collider experiment to the interaction cross section.

On the other hand, puzzling conflicts are seen to arise at the sub-atomic level when Newton's expectations are compared with today's Standard Model of particle physics. The Standard Model appears, against Newton's expectations, not to be self-similar, simple or purposeful. The last of these conclusions appears to the author to be the most serious.

The conflicts between the Standard Model and Newton's hypotheses may of course result from the presence of an error or errors in either the Standard Model or in the hypotheses, or in both. If Newton's hypotheses are valid, they may yet provide a useful guide to new physics at the subatomic level.

Some historical precursors to the Standard Model appear to be less conflicted by Newton's hypotheses, but they have not been developed to the extent that the Standard Model has, and there is no guarantee that they can be.

An observational programme has been described here which, if carried through, would test one possible alternative to the Standard Model that follows Newton's hypotheses. At the very least, it would supply an exquisitely detailed view of the proton. Other alternatives to the Standard Model that have not been considered here might fare better. Precursors to the Standard Model only were considered in the above.

**Acknowledgements**
The author acknowledges discussions with the late Bernard Feld, Kenneth Johnson, Gunnar Källén, Jun Sakurai and Julian Schwinger on the early stages of the work described above, with colleagues in astronomy more recently (Yock and Muraki 2017), and with many former students.